\documentclass[pdflatex,sn-mathphys-num]{sn-jnl}

\usepackage{graphicx}
\usepackage{amsmath,amssymb,amsfonts}
\usepackage{bm}
\usepackage{booktabs}
\usepackage{placeins}
\usepackage{soul}
\newcommand{\paircoh}{\Delta_{\mathrm{coh}}}
\newcommand{\Ract}{\mathcal{R}_{\mathrm{act}}}

\begin{document}

\title[Graph-structured VQE for fermionic pairing]{Graph-Structured Number-Conserving Variational Quantum Eigensolver for Fermionic Pairing Hamiltonians}

\author*[1]{\fnm{Abhishek} \sur{}}\email{abhishek@ph.iitr.ac.in}
\author[1]{\fnm{Nabeel} \sur{Salim}}\email{nabeel\_s@ph.iitr.ac.in}
\author[1,2]{\fnm{P.} \sur{Arumugam}}\email{arumugam@ph.iitr.ac.in}

\affil*[1]{\orgdiv{Department of Physics}, \orgname{Indian Institute of Technology Roorkee}, \orgaddress{\city{Roorkee}, \postcode{247667}, \state{Uttarakhand}, \country{India}}}
\affil[2]{\orgname{Centre for Photonics and Quantum Communication Technology}, \orgaddress{\orgname{Indian Institute of Technology Roorkee}, \city{Roorkee}, \postcode{247667}, \state{Uttarakhand}, \country{India}}}

\abstract{Simulating strongly correlated fermionic pairing in the presence of rotational and pair-breaking fields requires deep quantum circuits. We present a graph-structured variational quantum eigensolver whose pair-transfer and single-excitation rotations follow the nonzero pairing and one-body mixing edges of the Hamiltonian. The circuit conserves particle number exactly and uses one parameter per retained edge. We benchmark one layer against exact fixed-number diagonalization for 1500, $M=8$ Hamiltonians represented by 16 qubits. Its mean energy error rises from 0.86 keV without one-body driving to 691 keV at the strongest drive; the one-layer circuit loses accuracy as pair breaking strengthens. In a matched eight-qubit comparison, the graph circuit reaches the 0.42 keV high-drive error of a pair-plus-all-singles circuit with 8 instead of 34 parameters. Fixed-number Adaptive Derivative-Assembled Pseudo-Trotter VQE reaches 0.056 keV with 304 decomposed controlled-NOT gates and iterative pool screening, while a 52-parameter, single-repetition unitary coupled-cluster singles-and-doubles circuit gives 17.6 keV with 2752 such gates. At a separate twelve-qubit point, a second graph layer reduces the high-drive error from 378 to 35 keV. Across the exact grid, an off-diagonal pair-coherence scale tracks the leading pair-density eigenvalue, condensate fraction, and interaction energy. Cranked zirconium Hamiltonians provide the benchmark instances and tune the strength of one-body pair breaking.}

\keywords{variational quantum eigensolver, Fermionic pairing, particle-number conservation, graph-structured ansatz, quantum simulation, pair-density matrix}

\maketitle

\section{Introduction}\label{sec:introduction}

Pairing Hamiltonians are a recurring reduced description of correlated fermions in superconducting, molecular, confined, and nuclear systems~\cite{bcs1957,richardson1963,dukelsky2004,elfving2021seniority,ring1980}. In ultrasmall metallic grains, discrete single-particle spectra and magnetic fields expose finite-size pair breaking beyond bulk mean-field intuition~\cite{clogston1962,braun1999grains}. Their defining two-body process transfers a correlated pair between orbitals, while external fields or intrinsic one-body couplings can split or mix individual fermionic states and compete with pair coherence. Although reduced pairing models are simpler than general two-body Hamiltonians, their fixed-particle-number Hilbert spaces still grow combinatorially with the number of active orbitals.

The variational quantum eigensolver (VQE) offers a hybrid route to correlated eigenstates by preparing a parameterized state on a quantum register and minimizing its measured energy classically~\cite{peruzzo2014,cerezo2021,preskill2018, Pooja_2021, Nifeeya_2025, ashutosh2025, RRVQE}. Its accuracy and cost depend strongly on the ansatz. Hardware-efficient circuits are easy to deploy but may explore symmetry sectors and directions unrelated to the target Hamiltonian. General fermionic unitary coupled-cluster constructions are physically motivated but can introduce a large excitation pool~\cite{bartlett2007cc,romero2018ucc}. Symmetry-preserving circuits reduce this search space by enforcing conserved quantum numbers at the gate level~\cite{gard2020symmetry,anselmetti2021qnp,lacroix2023symmetry}. Pair-unitary coupled-cluster doubles and related paired constructions provide compact descriptions when pair transfer is dominant~\cite{elfving2021seniority,lee2019kupccgsd}.

We construct a graph-structured fixed-particle-number ansatz for pairing Hamiltonians with off-diagonal one-body orbital mixing. Pair-transfer rotations follow nonzero pairing edges, and single-excitation rotations follow nonzero one-body mixing edges, tying every parameter to a retained Hamiltonian coupling. Generators outside these graphs are excluded at one layer, although commutators and higher-rank correlations can still make them relevant to the exact state. Repeated graph layers enlarge the accessible manifold, whereas adaptive pool methods such as the Adaptive Derivative-Assembled Pseudo-Trotter VQE (ADAPT-VQE) can select additional directions explicitly~\cite{grimsley2019adapt}.

We test the method with cranked Nilsson-plus-pairing Hamiltonians. Quantum algorithms have also been applied to nuclear observables~\cite{singh2025linearresponse,pooja2022,GCM}.  Nuclear pairing has already motivated several quantum algorithms and circuit encodings~\cite{dumitrescu2018cloud,kiss2022li6,zhang2024nppairing,yoshida2026trapped,sarma2026lowdepth}. Zirconium nuclei near $A=80$ exhibit competing configurations and shape coexistence~\cite{lister1987,zheng2014zr}. We use $^{80,82,84}$Zr because deformation changes both the single-particle spectrum and the nonzero Coriolis mixing graph, while rotational frequency controls the strength of one-body pair breaking. The resulting grid tests the ansatz from pair-dominated to strongly mixed regimes. These nuclei provide algorithmic benchmark instances; the truncated Hamiltonian omits the core and macroscopic terms needed for precision shapes or spectroscopy.

We formulate the ansatz for a general graph-defined paired-fermion Hamiltonian and compare it under common conventions with pair-only, pair-plus-all-singles, number-preserving hardware-efficient, unrestricted hardware-efficient, unitary coupled-cluster singles-and-doubles (UCCSD), and fixed-number ADAPT-VQE baselines. Layer-depth and one-body-graph tests are benchmarked against exact fixed-number diagonalization. Complete exact validation covers the 1500-point sixteen-qubit grid, and the off-diagonal pairing-coherence scale is compared with established diagnostics of the pair-density matrix. Because this regime is classically tractable, exact references identify where the compact circuit succeeds and where it requires additional depth.

\section{Theoretical model and graph representation}\label{sec:model}
\subsection{Paired-fermion Hamiltonian with one-body mixing}
We consider a system of interacting fermions governed by strong two-body pairing and one-body orbital mixing. We define a basis of $M$ pair orbitals, where each orbital $k$ contains two time-reversed states, denoted $k+$ and $k-$. The creation and annihilation of a correlated pair in orbital $k$ are described by 
\begin{equation}
P_k^\dagger=a_{k+}^\dagger a_{k-}^\dagger,
\qquad
P_k=a_{k-}a_{k+}.
\end{equation}
For one or more external control parameters denoted collectively by $\lambda$, we study the fixed-particle-number sector of
\begin{equation}
\hat H(\lambda)=\sum_p \epsilon_p(\lambda)\hat n_p+\sum_{p<q}\left[t_{pq}(\lambda)a_p^\dagger a_q+\mathrm{H.c.}\right]-\sum_{kl}G_{kl}(\lambda)P_k^\dagger P_l.
\label{eq:generic_hamiltonian}
\end{equation}
Indices $p,q$ label the $2M$ individual spin orbitals, whereas $k,l$ label the $M$ partner pairs. The first term contains diagonal single-particle energies, the second mixes individual fermionic orbitals, and the last transfers correlated pairs. A diagonal field, such as a pure level splitting, changes $\epsilon_p$ but does not by itself generate a single-excitation edge. The ansatz developed below targets the off-diagonal mixing encoded by $t_{pq}$ together with the pairing interaction. General four-index interactions and complex hopping phases lie outside the present operator set.

For $N_{\mathrm{act}}$ active particles, exact diagonalization acts in a determinant space of dimension
\begin{equation}
D_N=\binom{2M}{N_{\mathrm{act}}}.
\label{eq:dimension}
\end{equation}
At half filling, $D_N=\binom{2M}{M}$. Under the Jordan--Wigner mapping~\cite{jordan1928}, the same spin-orbital register uses $2M$ qubits. The $2M$-qubit count describes representation size alone; end-to-end cost also includes state preparation, circuit depth, measurement, optimization, noise, and error correction.

\subsection{Cranked Nilsson-plus-pairing realization}

The general Hamiltonian identifies the relevant operator classes; the nuclear realization supplies a controlled family of their coefficients. The benchmark instantiates Eq.~\eqref{eq:generic_hamiltonian} in a deformation-dependent Nilsson basis~\cite{nilsson1955,ring1980}. For deformation $\delta$ and cranking energy $\hbar\omega$, the canonical active-space Routhian is
\begin{equation}
\hat{\mathcal R}_{\mathrm{act}}(\delta,\omega)=\sum_k\epsilon_k(\delta)(\hat n_{k+}+\hat n_{k-})-G\sum_{kl}P_k^\dagger P_l-(\hbar\omega)\frac{\hat J_x}{\hbar}+\lambda_{\mathrm p}(\hat N-N_{\mathrm{act}})^2.
\label{eq:routhian}
\end{equation}
Writing the drive as $(\hbar\omega)(\hat J_x/\hbar)$ makes both factors dimensionally explicit: $\hbar\omega$ is plotted in MeV and $\hat J_x/\hbar$ is dimensionless. The angular-momentum operator is
\begin{equation}
\hat J_x=\sum_{ij}(j_x)_{ij}a_i^\dagger a_j,
\end{equation}
with matrix elements reconstructed from the deformation-dependent Nilsson eigenvectors rather than inserted phenomenologically. At fixed deformation, the alignment of an exact eigenstate, or of a stationary optimized variational state, obeys the Hellmann--Feynman/envelope relation
\begin{equation}
\frac{\langle\hat J_x\rangle}{\hbar}
=-\left.\frac{\partial\mathcal R_{\min}(\delta,\omega)}{\partial(\hbar\omega)}\right|_{\delta}.
\label{eq:alignment}
\end{equation}
The number penalty vanishes identically for the symmetry-preserving ansatz and serves only as a numerical guard for alternative circuits.

The complete grid contains proton and neutron sectors of $^{80,82,84}$Zr, 25 deformation points in $-0.5\le\delta\le0.5$, and ten drive strengths in $0\le\hbar\omega\le1$ MeV. Each species is solved independently because the benchmark Hamiltonian contains no proton--neutron residual interaction. The production window has $M=8$ doubly degenerate orbitals, $N_{\mathrm{act}}=8$, and hence 16 qubits and $D_N=12870$ determinants per species. These 1500 points are a tensor-product parameter grid, not 1500 statistically independent physical systems. Active-window selection and coupling calibration are specified in Appendix~\ref{app:nuclear}.

\subsection{Hamiltonian support graphs}

The Routhian specifies the benchmark instances, but the circuit construction requires only the support of its off-diagonal couplings.  Let $V$ denote the set of vertices corresponding to the available physical orbitals in the chosen active space. Two graphs isolate these operators,
\begin{align}
\mathcal G_1&=(V_1,E_1), & E_1&=\{\{p,q\}:p<q,\ t_{pq}\ne0\},\\
\mathcal G_2&=(V_2,E_2), & E_2&=\{\{k,l\}:k<l,\ G_{kl}\ne0\}.
\end{align}
Vertices of $\mathcal G_1$ are individual spin orbitals, while vertices of $\mathcal G_2$ are pair orbitals. An edge of $\mathcal G_1$ denotes an allowed one-body mixing process; an edge of $\mathcal G_2$ denotes an allowed pair transfer. Diagonal one-body and pair-occupancy terms affect the energy but do not define excitation edges. In the constant-$G$ nuclear benchmark, $\mathcal G_1$ is selected from the nonzero $j_x$ matrix elements and $\mathcal G_2$ is complete.

\section{Graph-structured fixed-number VQE}\label{sec:method}

\subsection{Ansatz construction}
\begin{figure}[!htbp]
\centering
\includegraphics[width=\textwidth]{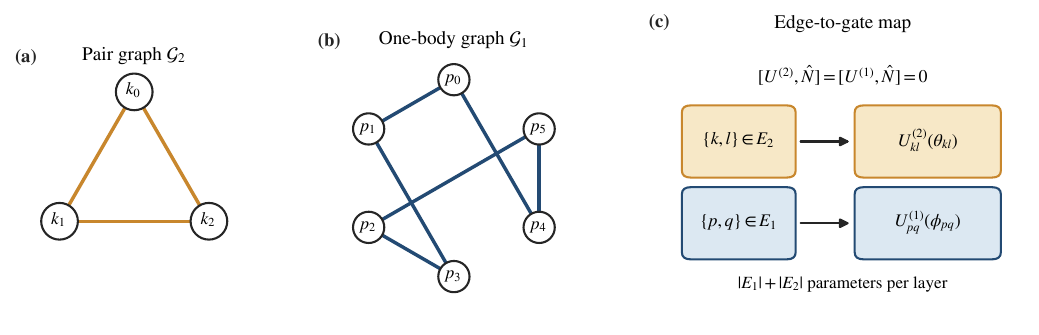}
\caption{Support graphs for three pair orbitals and six spin orbitals. (a) The complete three-vertex pair graph $\mathcal G_2$ defines three pair-transfer gates. (b) Each retained edge of the one-body graph $\mathcal G_1$ defines one single-excitation gate. (c) The edge-to-gate map assigns one variational parameter to every retained edge; both gate families conserve particle number.}
\label{fig:graph_concept}
\end{figure}

We construct a fixed-particle-number ansatz by placing variational gates strictly along the edges of the physical support graphs. Starting from a reference determinant $\vert{}\Phi_N\rangle$, one layer of the ansatz is written as
\begin{equation}
U_{\mathrm{layer}}(\bm\phi,\bm\theta)=
\prod_{(p,q)\in E_1}U^{(1)}_{pq}(\phi_{pq})
\prod_{(k,l)\in E_2}U^{(2)}_{kl}(\theta_{kl}),
\label{eq:layer}
\end{equation}
with
\begin{align}
U^{(1)}_{pq}(\phi)&=\exp\!\left[\frac{\phi}{2}(a_p^\dagger a_q-a_q^\dagger a_p)\right],\\
U^{(2)}_{kl}(\theta)&=\exp\!\left[\frac{\theta}{2}(P_k^\dagger P_l-P_l^\dagger P_k)\right].
\end{align}
The superscripts identify operator rank: $U^{(1)}$ is a one-body rotation, whereas $U^{(2)}$ transfers a correlated pair. The complete block is denoted $U_{\mathrm{layer}}$ to distinguish it from either gate family. Both generators commute with total particle number, so every circuit parameter preserves $N$ exactly. Repeating Eq.~\eqref{eq:layer} $L$ times gives $L(|E_1|+|E_2|)$ parameters. The pairing and one-body sectors generally do not commute. Products of repeated layers can therefore access additional commutator-generated directions, but finite layer count does not guarantee completeness or exact recovery. Figure~\ref{fig:graph_concept} defines the two support graphs and their gate correspondence. In the $M=8$ benchmark, $|E_1|=14$ and $|E_2|=28$, giving 42 parameters per layer and species.

Pair-unitary coupled-cluster doubles and quantum-number-preserving pair-exchange circuits use related generators~\cite{elfving2021seniority,anselmetti2021qnp}; here, the target Hamiltonian determines the excitation graph. Relative to generalized fermionic singles and doubles, this choice reduces the parameter pool from a generic polynomial set to the actual pair and one-body edges. The depth benchmark measures the accuracy gained by repeating the graph layer, while ADAPT-VQE tests explicit operator-pool enlargement. Neither construction is assumed to span the exact state at finite depth.

\subsection{Pairing observables at fixed particle number}

The number-conserving circuit fixes the symmetry sector, which also determines how pairing must be diagnosed. In particular, anomalous one-pair expectation values cannot serve as fixed-number observables.

For a number-conserving state, $\langle P_k\rangle=0$ because $P_k$ changes particle number by two. The conventional anomalous Bardeen--Cooper--Schrieffer (BCS) gap therefore vanishes by symmetry even when the state is strongly pair correlated~\cite{bcs1957,ring1980}. Pairing information instead resides in the two-body pair-density matrix
\begin{equation}
\rho_{kl}=\langle P_k^\dagger P_l\rangle.
\label{eq:pair_density}
\end{equation}
Yang's off-diagonal-long-range-order criterion connects pairing coherence to a macroscopically large eigenvalue of the two-body density matrix~\cite{yangpairing}. For the finite active spaces considered here, the primary diagnostics are the largest eigenvalue $\lambda_{\max}(\rho)$, the condensate fraction
\begin{equation}
f_{\mathrm{pair}}=\frac{\lambda_{\max}(\rho)}{N_{\mathrm{act}}/2},
\end{equation}
and the pairing interaction energy
\begin{equation}
E_{\mathrm{pair}}=-\sum_{kl}G_{kl}\rho_{kl}.
\end{equation}
For compact visualization we also use
\begin{equation}
\paircoh=G\sqrt{\sum_{k\ne l}|\rho_{kl}|},
\label{eq:paircoh}
\end{equation}
for the constant-$G$ benchmark. This basis-dependent, unnormalized quantity is a dimensional off-diagonal $l_1$ coherence scale. The BCS gap, a thermodynamic order parameter, and the Yang eigenvalue remain distinct diagnostics. Cross-window comparisons therefore use
\begin{equation}
C_{\mathrm{pair}}^{(l_1)}=
\frac{1}{M(M-1)}\sum_{k\ne l}|\rho_{kl}|.
\label{eq:normalized_coherence}
\end{equation}
The square-root form in Eq.~\eqref{eq:paircoh} gives the proxy the same energy scale as a pairing gap; alternative norms produce similar qualitative trends. Termwise magnitudes remove phase cancellation and therefore distinguish $\paircoh$ from $E_{\mathrm{pair}}$. Appendix~\ref{app:pair_impl} gives the statevector extraction and numerical Hermitization procedure.

\section{Computational assessment}\label{sec:assessment}

With the Hamiltonian, circuit manifold, and fixed-number diagnostics specified, we assess optimization behavior, accuracy against exact diagonalization, and logical circuit resources under common numerical conventions.

The VQE objective is evaluated with statevector simulation. Parameters are optimized with the limited-memory Broyden--Fletcher--Goldfarb--Shanno algorithm with bound constraints (L-BFGS-B)~\cite{zhu1997lbfgsb}. At fixed deformation, optimized parameters are warm-started from one control-parameter value to the next. Exact comparison is performed by projecting the Jordan--Wigner Hamiltonian into the $N_{\mathrm{act}}=8$ determinant sector and solving for its lowest eigenpair with a sparse Hermitian eigensolver.

Gate counts are reported after logical decomposition and exclude hardware transpilation. In the matched ansatz comparison, every circuit is decomposed to the same $\{R_z,\sqrt{X},X,\mathrm{CX}\}$ basis with an all-to-all logical connectivity assumption. Pair-only, graph-structured, pair-plus-all-singles, number-preserving hardware-efficient, and EfficientSU2 circuits use three initializations and identical L-BFGS-B settings. The number-preserving hardware-efficient circuit consists of a linear layer of excitation-preserving iSWAP-mode rotations. The unrestricted EfficientSU2 circuit is retained only as a symmetry-breaking control. Product-form UCCSD includes all occupied-to-virtual singles and doubles from the same reference; because of its synthesis cost, it is evaluated from one initialization. Fixed-number ADAPT-VQE uses a pool containing all number-conserving spin-orbital singles and all pair transfers, rather than generalized fermionic doubles. The pool contains 34 operators for $M=4$ and 81 for $M=6$. ADAPT starts from the common determinant and reports both inner objective evaluations and operator-pool gradient evaluations. We also evaluate, $\sigma_N^2=\langle(\hat N-N_{\mathrm{act}})^2\rangle$, to distinguish target-sector states from low-energy states obtained by changing particle number. We use noiseless statevector objectives to isolate errors from graph selection and circuit depth. Device noise and finite-shot uncertainty are excluded.

\subsection{Circuit realization}

\begin{figure}[!htbp]
\centering
\includegraphics[width=0.96\textwidth]{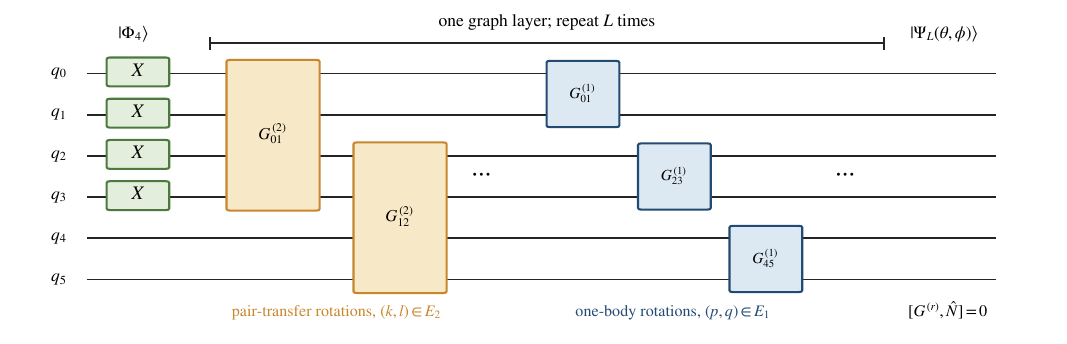}
\caption{Six-qubit Givens-rotation schematic for three pair orbitals and $N_{\mathrm{act}}=4$. The $X$ gates prepare the fixed-number reference. The four-wire $G^{(2)}$ gates represent pair-transfer rotations on edges of $E_2$, and the two-wire $G^{(1)}$ gates represent one-body rotations on edges of $E_1$. The ellipses stand for the remaining retained edges in the fixed application order. The complete graph layer is repeated $L$ times; internal gate decompositions are suppressed.}
\label{fig:ansatz_circuit}
\end{figure}

Figure~\ref{fig:ansatz_circuit} shows the circuit organization for an illustrative six-qubit, four-particle reference. The code first prepares the paired determinant, applies all pair-transfer gates in the fixed order of $E_2$, and then applies all one-body gates in the fixed order of $E_1$. In the schematic, $G^{(2)}$ and $G^{(1)}$ denote the Givens-style circuit realizations of $U^{(2)}$ and $U^{(1)}$, respectively. Each displayed gate represents one retained graph edge; the production layer contains every edge in $E_2$ and $E_1$. Repeating this sequence produces the $L$-layer ansatz. Internal two- and four-qubit decompositions are suppressed.

\subsection{Matched ansatz comparison at \texorpdfstring{$M=4$}{M=4}}

\begin{figure}[!htbp]
\centering
\includegraphics[width=0.95\textwidth]{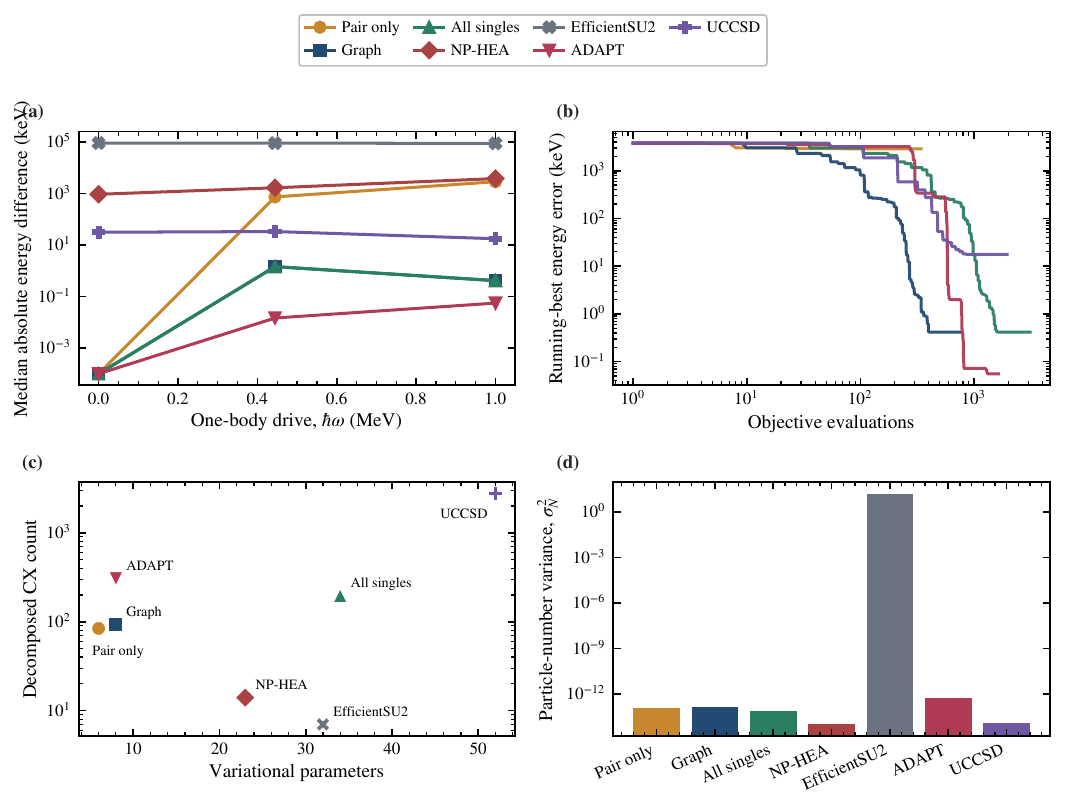}
\caption{Matched $M=4$, $N_{\mathrm{act}}=4$ ansatz comparison for the $^{80}$Zr proton instance at $\delta=+0.0417$ and three one-body drive strengths. (a) Median absolute energy difference from exact fixed-number diagonalization for the three-initialization circuits; ADAPT-VQE and UCCSD use one run. The EfficientSU2 value is not a fixed-$N$ variational error because that circuit leaves the target sector. (b) High-drive running-best traces for target-sector methods. (c) Parameter--CX tradeoff. (d) Particle-number variance $\sigma_N^2=\langle(\hat N-N_{\mathrm{act}})^2\rangle$ at the strongest drive. NP-HEA denotes the number-preserving hardware-efficient ansatz.}
\label{fig:matched_ansatz}
\end{figure}

Figure~\ref{fig:matched_ansatz} compares the principal alternatives under common statevector, reference-state, and circuit-decomposition conventions for $M=4$, $N_{\mathrm{act}}=4$. The pair-only and all-singles circuits isolate the two graph choices. At $\hbar\omega=1$ MeV, the graph-structured circuit reaches a 0.415 keV energy error with 8 parameters and 92 decomposed CX gates. Adding every number-conserving single-excitation edge gives the same error with 34 parameters and 196 CX gates. The pair-only circuit remains 2.872 MeV above exact, so one-body generators are needed when the drive breaks pairs.

Fixed-number ADAPT-VQE selects 8 operators from a 34-operator pool and lowers the endpoint error to 0.056 keV. Its evolved-operator circuit uses 304 decomposed CX gates, the inner optimizations require 1642 objective evaluations, and pool screening requires 306 gradient-component evaluations. The graph circuit requires 783 objective evaluations at the same point. ADAPT reaches the lower error, whereas the graph construction has lower decomposed depth and avoids iterative pool screening.

The one-layer number-preserving hardware-efficient ansatz has 23 parameters and 14 CX gates but remains 3.806 MeV above exact at the endpoint. The unrestricted EfficientSU2 control converges outside the target sector. Figure~\ref{fig:matched_ansatz}(a) therefore reports only the magnitude of its energy difference; panel (d) shows that this value cannot be interpreted as a fixed-$N$ variational error. The single-start, single-repetition UCCSD errors are 31.4, 33.4, and 17.6 keV over the three drive strengths. At the endpoint, this tested ordering requires 1961 objective evaluations, 52 parameters, and 2752 decomposed CX gates. The comparison is limited to these implementations and optimization settings.

\subsection{Depth--accuracy tradeoff at \texorpdfstring{$M=6$}{M=6}}

At the strongly driven $^{80}$Zr proton point $\delta=-0.25$ and $\hbar\omega=1$ MeV, the $M=6$ fixed-number dimension is 924. Here the complete pair graph contributes 15 parameters and the local one-body graph contains six edges, giving 21 parameters and 234 decomposed CX gates in one layer. The one-layer energy lies 378.4 keV above exact. A second layer is initialized by embedding the optimized first-layer parameters and then reoptimized. It reduces the energy error to 34.8 keV; the alignment error changes from $-16.4\%$ to $-0.07\%$, while the magnitude of the pair-coherence error remains below $2.1\%$. The second layer doubles the parameter count to 42 and the decomposed CX count to 468.

\begin{figure}[!htbp]
\centering
\includegraphics[width=0.82\textwidth]{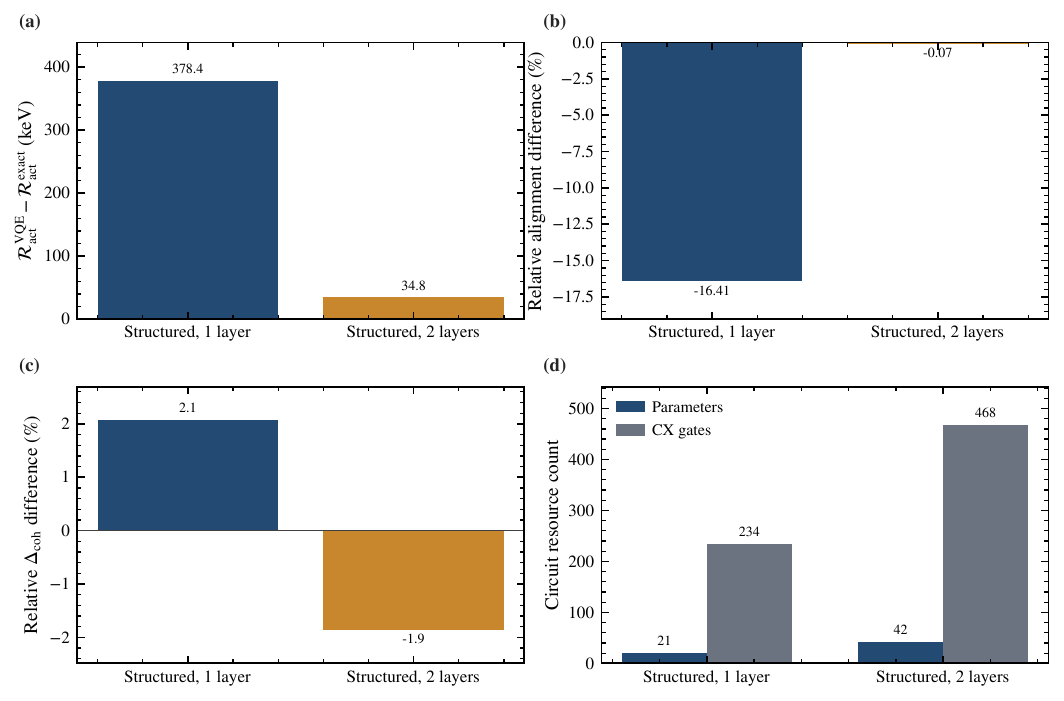}
\caption{Exact fixed-number depth benchmark at $M=6$, $\delta=-0.25$, and $\hbar\omega=1$ MeV. Panels show (a) variational energy error, (b) relative alignment difference, (c) relative pair-coherence difference, and (d) parameter and decomposed CX counts. The warm-started second layer improves energy and alignment at increased circuit cost.}
\label{fig:depth}
\end{figure}

At this Hamiltonian point, the second layer adds expressivity; one point cannot establish monotonic convergence with layer count. A shallow 48-parameter EfficientSU2 calculation at the same point has a 10.5 MeV error and is retained as one illustrative hardware-efficient control.

\subsection{One-body graph ablation at \texorpdfstring{$M=6$}{M=6}}

\begin{figure}[!htbp]
\centering
\includegraphics[width=0.82\textwidth]{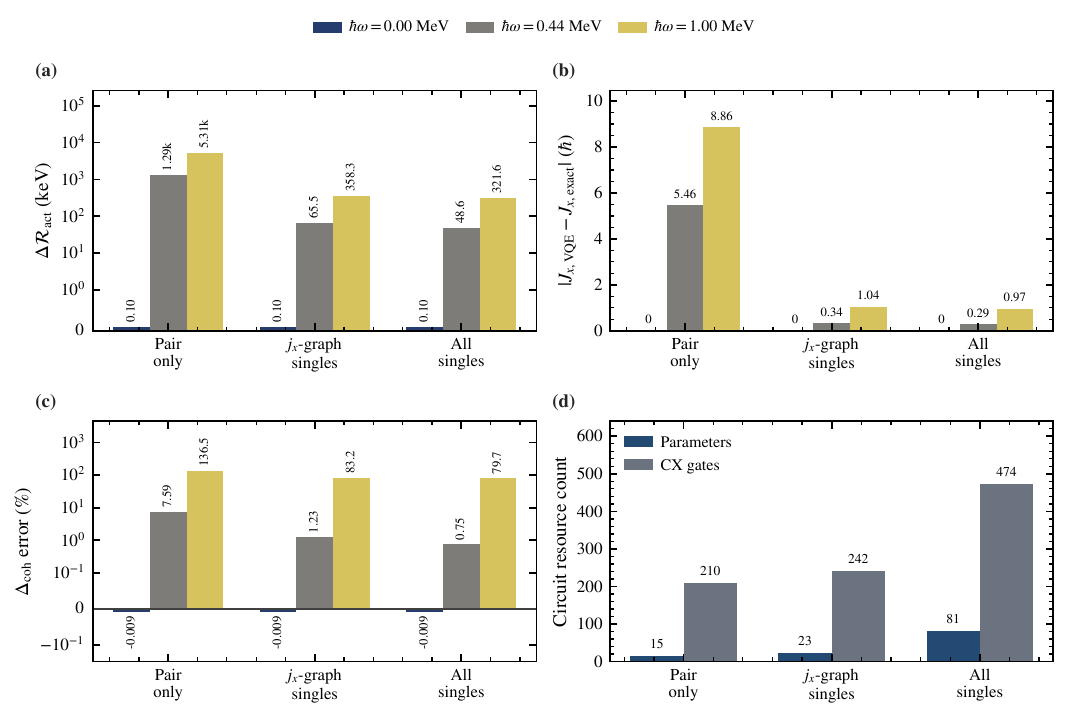}
\caption{One-body graph ablation for a one-layer $M=6$ circuit at $\delta=+0.0417$. Panels show (a) energy error, (b) absolute alignment difference, (c) relative pair-coherence error, and (d) parameter and decomposed CX counts. Panels (a) and (c) use symmetric-log scales to retain weak-drive values; panel (b) uses a zero-based linear scale.}
\label{fig:ablation}
\end{figure}

The graph-ablation calculation uses a different $^{80}$Zr proton point, $\delta=+0.0417$. Its one-body graph contains eight rather than six edges, so the structured circuit has $15+8=23$ parameters and 242 decomposed CX gates. Figure~\ref{fig:ablation} compares pair transfer only, pair transfer plus these eight nonzero one-body edges, and pair transfer plus all 66 number-conserving single edges. Without one-body driving, all variants give the same 0.102 keV error. At an intermediate drive, selected singles reduce the pair-only error from 1286 to 65 keV, while all singles reach 49 keV. At the strongest drive, the corresponding errors are 5314, 358, and 322 keV. The selected graph therefore retains most of the all-singles improvement with 23 rather than 81 parameters and 242 rather than 474 decomposed CX gates.

The similar residual high-drive errors for selected and all singles indicate that omitted one-body edges are not the dominant limitation at this point; additional depth or other generators are more relevant candidates. Appendix~\ref{app:matched_m6} gives the matched $M=6$, $\delta=+0.0417$ comparison. At the strongest drive, graph restriction gives 358 keV with 23 parameters and 242 CX gates, compared with 322 keV, 81 parameters, and 474 CX gates for all singles. A 30-operator ADAPT circuit lowers the finite-budget error to 128 keV but uses 1000 CX gates, 32870 objective evaluations, and 2430 pool-gradient evaluations. Its final pool gradient remains above threshold, so this is an iteration-capped result rather than a converged ADAPT optimum.

\subsection{Complete sixteen-qubit grid validation}

\begin{figure}[!htbp]
\centering
\includegraphics[width=0.95\textwidth]{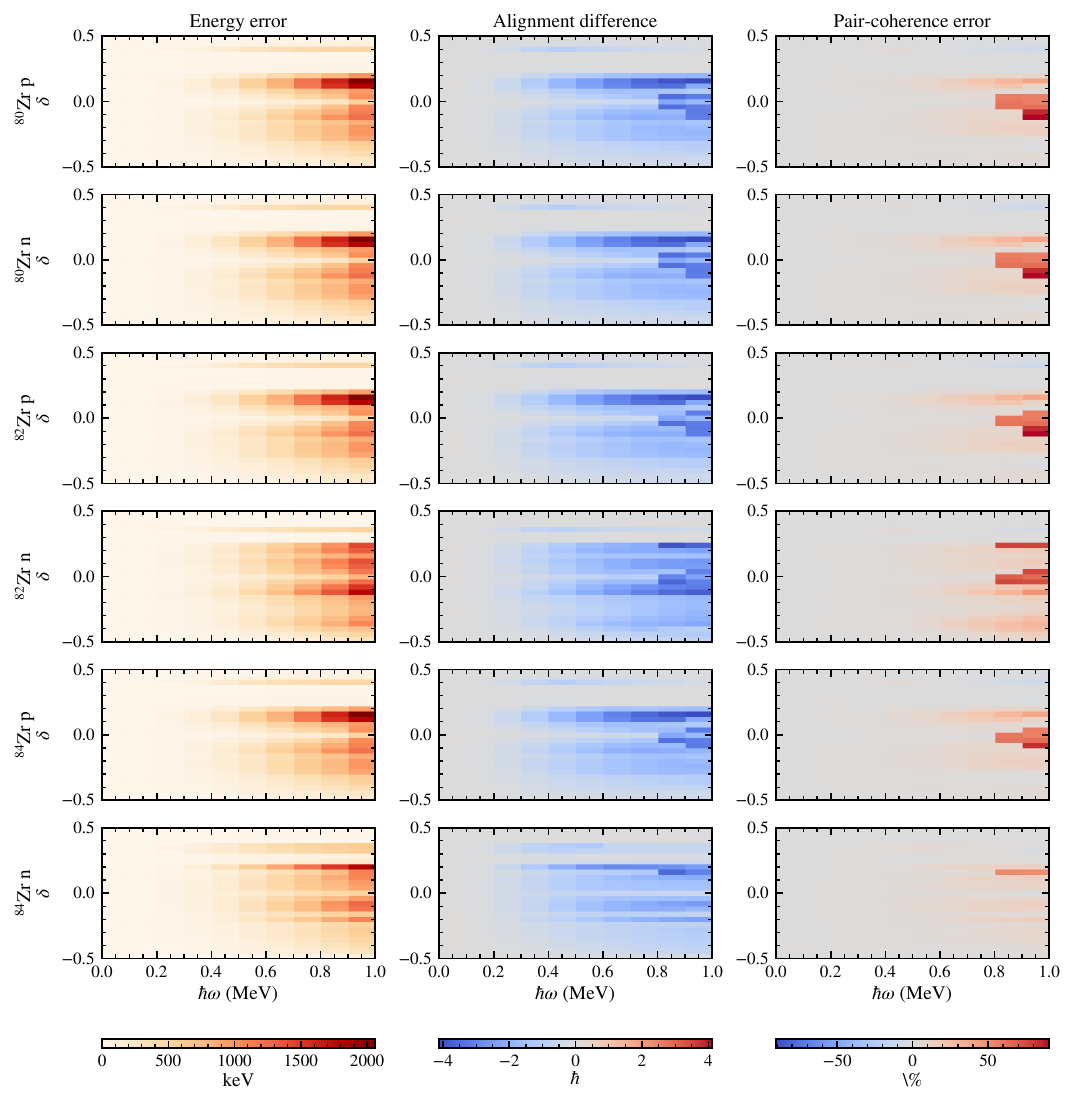}
\caption{Complete exact fixed-number error maps for the one-layer $M=8$ circuit. Rows identify isotope and particle species; columns show active-space energy error, signed alignment difference, and relative pair-coherence error over the deformation--drive grid. Variational bias grows continuously as one-body pair breaking increases.}
\label{fig:complete}
\end{figure}

We next diagonalize the complete $M=8$ tensor-product grid: three isotopes, two species, 25 deformation values, and ten drive strengths, giving 1500 exact fixed-number Hamiltonian instances. Table~\ref{tab:complete} and Fig.~\ref{fig:complete} show the systematic loss of one-layer accuracy as one-body pair breaking increases. The mean energy error rises from 0.864 keV without driving to 108 keV at $\hbar\omega=0.444$ MeV and 691 keV at 1 MeV. The mean absolute alignment error reaches $1.37\hbar$, and the mean relative $\paircoh$ error reaches $17.2\%$ at the endpoint. Absolute energy errors are reported because normalization by the large total active-space Routhian would obscure the variational bias.

\begin{center}
\refstepcounter{table}\label{tab:complete}
\begin{minipage}{0.66\textwidth}
{\footnotesize\textbf{Table \thetable}\quad Complete-grid exact validation of the one-layer $M=8$ VQE. Each frequency column contains 150 isotope--species--deformation points.\par\medskip
\centering
\begin{tabular}{@{}lccc@{}}
\toprule
Quantity & \multicolumn{3}{c}{$\hbar\omega$ (MeV)}\\
& 0.000 & 0.444 & 1.000\\
\midrule
Mean $\Delta\Ract$ (keV) & 0.864 & 108.394 & 690.679\\
Maximum $\Delta\Ract$ (keV) & 3.320 & 311.175 & 2061.093\\
Mean $|\Delta J_x|$ ($\hbar$) & 0.000 & 0.644 & 1.374\\
Mean relative $\paircoh$ error (\%) & 0.026 & 1.538 & 17.157\\
\bottomrule
\end{tabular}
\par}
\end{minipage}
\end{center}

For the fixed-number one-layer $M=8$ validation, all energy differences are nonnegative, as required by the variational principle. Their drive dependence is much larger than the representative multistart spread reported in Appendix~\ref{app:optimizer}, consistent with depth-limited expressivity at high drive. The multistart test covers one representative point, so optimizer failures elsewhere and barren plateaus over the full grid remain untested.

\FloatBarrier
\section{Validation of fixed-number pairing diagnostics}\label{sec:pair_validation}

\begin{figure}[!htbp]
\centering
\includegraphics[width=0.88\textwidth]{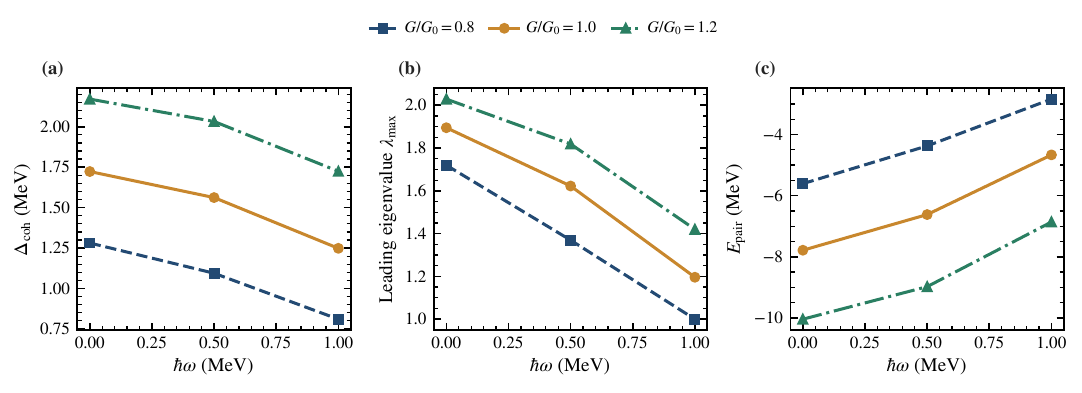}
\caption{Exact fixed-number pairing diagnostics under interaction-strength variation. Panels show (a) $\paircoh$, (b) the leading pair-density eigenvalue, and (c) pairing energy for $G/G_0=0.8,1.0,1.2$. The common legend applies to all panels.}
\label{fig:g_sensitivity}
\end{figure}

Figure~\ref{fig:g_sensitivity} tests the pair-density diagnostics under a $\pm20\%$ change of the interaction strength at one representative $M=8$ point. Increasing the one-body drive suppresses $\paircoh$, the leading pair-density eigenvalue, and the magnitude of the pairing energy together. Increasing $G$ raises all three measures. The calculation measures the sensitivity of a finite-system coherence proxy; the Yang eigenvalue, a broken-symmetry gap, and a thermodynamic transition remain distinct quantities.

\begin{figure}[!htbp]
\centering
\includegraphics[width=\textwidth]{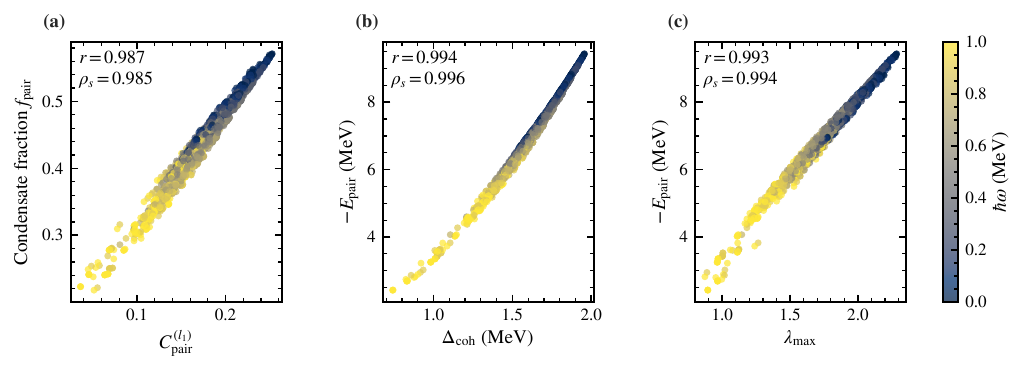}
\caption{Pair-density diagnostics over the complete exact $M=8$ grid. Panels compare (a) normalized $l_1$ coherence with condensate fraction, (b) $\paircoh$ with pairing energy, and (c) the leading pair-density eigenvalue with pairing energy. Color denotes $\hbar\omega$. The quantities are highly correlated but remain distinct.}
\label{fig:pair_correlations}
\end{figure}

Across all 1500 exact states, the normalized off-diagonal $l_1$ coherence and condensate fraction have Pearson and Spearman correlations of 0.987 and 0.985, respectively. The dimensional $\paircoh$ scale and $-E_{\mathrm{pair}}$ have correlations of 0.994 and 0.996. The correlations are high but below unity, so the observables remain numerically distinct (Fig.~\ref{fig:pair_correlations}).

The leading eigenvalue, condensate fraction, and interaction energy remain the formally grounded diagnostics. Within the finite active space, the smooth decrease measures suppressed pair coherence; it cannot resolve a sharp thermodynamic transition or complete pairing collapse. On shot-based hardware, termwise magnitudes and the square root in Eq.~\eqref{eq:paircoh} will introduce nonlinear finite-sampling bias; measurement grouping and uncertainty mitigation are left for future work.

\FloatBarrier
\section{Cranked zirconium as an application benchmark}\label{sec:application}

\begin{figure}[!htbp]
\centering
\includegraphics[width=0.88\textwidth]{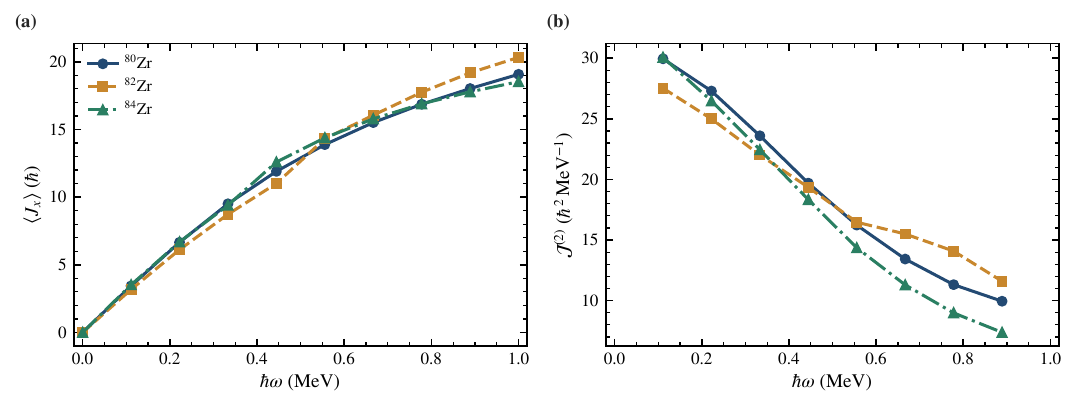}
\caption{Application-level response of the one-layer circuit for the three zirconium benchmark sets. (a) Alignment $\langle J_x\rangle$ and (b) centered finite-difference dynamical moment of inertia $\mathcal J^{(2)}$. The endpoint frequencies are omitted from panel (b).}
\label{fig:rotation}
\end{figure}

The zirconium benchmark spans deformation-dependent spectra and Coriolis graphs across three neighboring isotopes. Increasing frequency drives each instance from a pair-dominated regime toward broken-pair configurations. Figure~\ref{fig:rotation} shows the compact-circuit response along selected active-space grid paths. Time-reversal symmetry gives $J_x=0$ at zero drive. The dynamical moment of inertia is evaluated as $\mathcal J^{(2)}=d\langle J_x\rangle/d\omega=\hbar\,d\langle J_x\rangle/d(\hbar\omega)$ and is reported in $\hbar^2\,\mathrm{MeV}^{-1}$. Endpoint values are omitted because the centered finite-difference stencil is unavailable there.

The complete exact benchmark shows increasing one-layer bias in the same high-drive region. The one-body mixing that enlarges the relevant configuration manifold also marks where a second layer, an adaptive extension, or additional excitation rank becomes necessary.

The nuclear inputs contain only the truncated active-space Hamiltonian. The active window changes at some deformation boundaries, and inactive-orbital, core, and macroscopic deformation energies are absent. These omissions limit nuclear-shape interpretation. The resulting sparse, parameter-dependent fermionic Hamiltonians serve as a benchmark set with exact reference solutions; Appendix~\ref{app:surfaces} documents representative truncated surfaces.

\section{Conclusion}\label{sec:conclusion}

Hamiltonian-edge selection gives a compact, exactly number-conserving circuit for paired fermions under a one-body mixing field. At $M=4$, the graph circuit reaches the 0.415 keV high-drive error of the all-singles circuit with 8 rather than 34 parameters and 92 rather than 196 CX gates. At $M=6$, the selected one-body graph retains most of the all-singles accuracy with 23 rather than 81 parameters. Fixed-number ADAPT-VQE reaches smaller small-space energy errors but requires a larger evolved-operator circuit and repeated pool-gradient screening. The tested one-repetition UCCSD circuit is deeper and less accurate at $M=4$, although its single initialization does not characterize UCCSD in general.

Across 1500 sixteen-qubit instances, one layer is nearly exact without one-body driving and develops a growing variational bias as pair breaking strengthens. A second layer reduces a separate twelve-qubit high-drive error from 378 to 35 keV. At that point, added depth recovers part of the missing expressivity. The off-diagonal coherence proxy follows the leading pair-density eigenvalue, condensate fraction, and interaction energy across the exact grid. Its basis dependence and nonlinear finite-shot bias still require pairing it with the pair-density spectrum or interaction energy.

The construction also applies to nonuniform pairing matrices and to tunneling, orbital-mixing, or spin-mixing fields: the one-body terms change $E_1$, while the pairing matrix changes $E_2$, without changing the circuit rule. General four-index interactions or correlations outside the pairing channel require additional generators or adaptive selection. The present statevector calculations measure ansatz error and logical resources, not runtime advantage. Tests beyond the exact-validation regime must include compiled depth, Pauli-term grouping, shot cost, optimizer cost, and noise.

\appendix

\section{Nuclear benchmark specification}\label{app:nuclear}

At each deformation, the $M$ Nilsson orbitals nearest the auxiliary Fermi level are selected by energy. The stored dimensionless single-particle energies are rescaled with the isotope-dependent oscillator energy, and the same eigenvectors are used to reconstruct $j_x$. The deformation mesh contains 25 points and the frequency mesh contains ten points. An overlap audit matches neighboring Nilsson eigenvectors within symmetry blocks; the selected $M=8$ boundary changes at 13--15 of 24 adjacent deformation steps, depending on isotope and species. These frequent boundary changes prevent interpreting the deformation minima as equilibrium nuclear shapes.

The constant pairing strength is $G=0.5202$ MeV for both species and the numerical number penalty is $\lambda_{\mathrm p}=5$ MeV. The value of $G$ was calibrated in the $^{84}$Zr, $M=8$, $\delta=0$ window so that the mean proton and neutron classical BCS gaps reproduce a reference 1.878 MeV empirical scale from the National Nuclear Data Center NuDat 3 database~\cite{nudat3}. It is an effective active-window coupling, not a universal microscopic constant. The $\pm20\%$ calculation in Fig.~\ref{fig:g_sensitivity} quantifies the sensitivity of the pairing diagnostics to this choice.

The active-space quantity in Eq.~\eqref{eq:routhian} is canonical: no deformation-dependent chemical-potential subtraction is made. In a fixed-number sector, any Fermi-level term proportional to $\hat N$ is a scalar; retaining a deformation-dependent scalar would corrupt comparisons across the parameter mesh. The auxiliary Fermi level is used only to select the active window and its reference occupations.

\section{Pair-density implementation}\label{app:pair_impl}

Each $P_k^\dagger P_l$ operator is mapped to the Jordan--Wigner register and evaluated directly on the statevector. Diagonal terms are omitted from Eqs.~\eqref{eq:paircoh} and \eqref{eq:normalized_coherence}, leaving inter-orbital pair coherence rather than pair occupancy. The full matrix is Hermitized numerically before evaluating $\lambda_{\max}$ and $f_{\mathrm{pair}}$. The pairing energy retains complex phases and signs before its real expectation value is taken.

For $M=8$, Hermiticity leaves 28 independent off-diagonal pair correlators. A hardware implementation would decompose these into Pauli strings and group compatible measurements. Because absolute values and square roots are nonlinear, unbiased estimates of individual correlators do not produce an unbiased estimate of $\paircoh$; bootstrap or likelihood-based uncertainty propagation would therefore be required for finite-shot data.

\FloatBarrier
\section{Truncated parameter surfaces}\label{app:surfaces}

\begin{figure}[!htbp]
\centering
\includegraphics[width=0.88\textwidth]{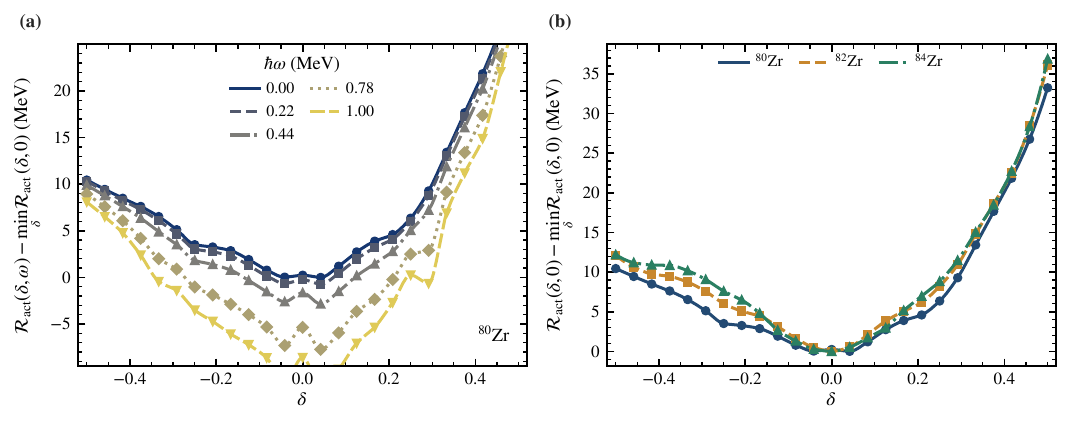}
\caption{Representative canonical truncated active-space energy surfaces. (a) Five selected drive strengths for the $^{80}$Zr benchmark, all referenced to the zero-drive minimum. (b) The three isotope families without driving, each referenced to its own minimum. These are benchmark Hamiltonian surfaces, not total nuclear Routhians or shape predictions.}
\label{fig:surfaces}
\end{figure}

Figure~\ref{fig:surfaces} documents representative Hamiltonians from the parameterized benchmark set. The curves omit inactive, core, and macroscopic deformation contributions and show the inputs used by the algorithm.

\FloatBarrier
\section{Optimizer diagnostics}\label{app:optimizer}

\begin{figure}[!htbp]
\centering
\includegraphics[width=0.72\textwidth]{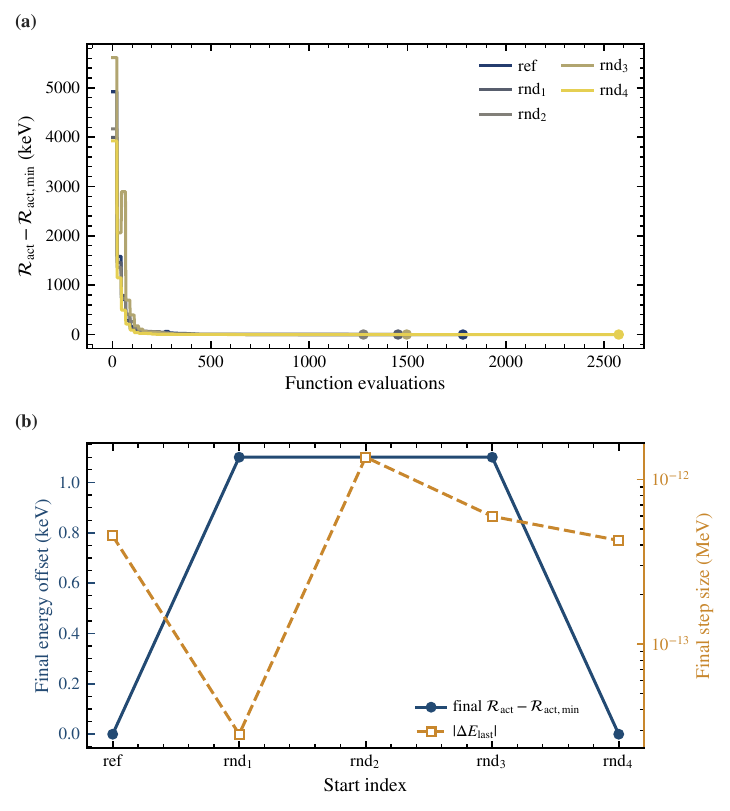}
\caption{Representative VQE optimizer diagnostics for the $^{80}$Zr proton sector at $M=6$, $\delta=-0.25$, and $\hbar\omega=1$ MeV. (a) Energy traces for five initializations. (b) Final offsets from the best run and the final callback step size. The optimizer is reproducible even though a one-layer ansatz retains a finite exact-benchmark error.}
\label{fig:optimizer}
\end{figure}

Figure~\ref{fig:optimizer} reports five initializations at the representative $M=6$ high-drive point. They converge to a 1.100 keV final-energy band, with final callback energy steps at or below the $10^{-12}$ MeV scale. The narrow spread shows local optimizer reproducibility. Variational completeness remains a separate issue: the best one-layer result lies 378 keV above exact, while the two-layer result reduces this to 35 keV.

Representative $M=8$ endpoint calculations require 1968--7995 function evaluations and 33--112 L-BFGS-B iterations, depending on isotope and species. Their final callback step sizes remain at the $10^{-12}$ MeV scale.

\FloatBarrier
\section{Extended matched comparison at \texorpdfstring{$M=6$}{M=6}}\label{app:matched_m6}

\begin{figure}[!htbp]
\centering
\includegraphics[width=0.95\textwidth]{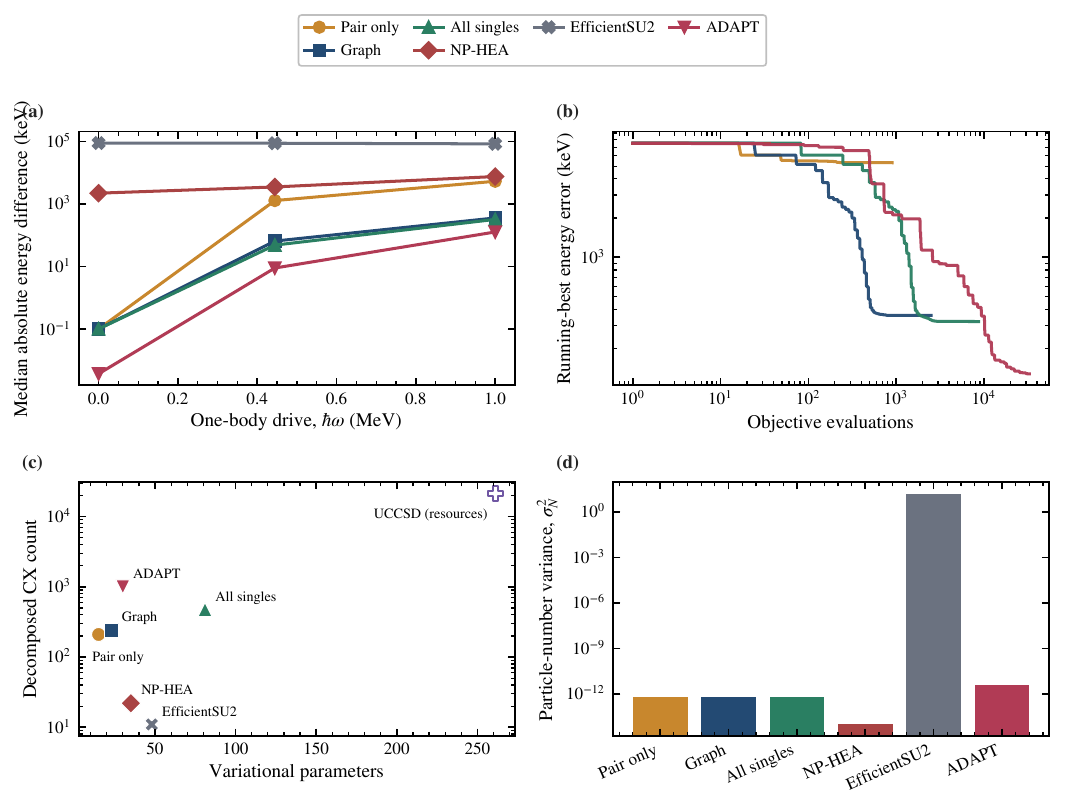}
\caption{Extended matched comparison for the $^{80}$Zr proton instance with $M=6$, $N_{\mathrm{act}}=6$, and $\delta=+0.0417$. (a) Absolute energy differences at three drive strengths, with ADAPT represented by the 30-operator run. (b) High-drive running-best objective traces. (c) Parameter--CX resources; UCCSD is resource-only at this size. (d) Target-sector error. The finite ADAPT errors at intermediate and high drive are iteration-capped values, not converged optima.}
\label{fig:matched_m6}
\end{figure}

Figure~\ref{fig:matched_m6} repeats the matched comparison in the 12-qubit, 924-dimensional fixed-number sector. The ADAPT calculation uses a 30-operator cap. It reaches its gradient threshold without driving, while the intermediate- and high-drive calculations terminate at the cap with final maximum gradients of 0.032 and 0.049 MeV, respectively. At $M=6$, UCCSD contributes a resource point: its 261-parameter, 21264-CX logical circuit was not optimized, so no accuracy value is reported.

At the high-drive endpoint, the capped ADAPT result has relative alignment and pair-coherence errors of $-8.9\%$ and $78.0\%$, respectively; the corresponding graph-circuit errors are $-11.8\%$ and $83.2\%$. The lower ADAPT energy error is therefore accompanied by sizeable observable errors. The $M=6$ comparison reports a finite-budget accuracy--resource tradeoff.

\FloatBarrier
\backmatter

\noindent\textbf{Acknowledgements.}\ 
The authors acknowledge institutional and computational support from the Department of Physics, Indian Institute of Technology Roorkee.

\section*{Declarations}

\noindent\textbf{Funding.}\ 
This work was supported by the Science and Engineering Research Board under Grant No. CRG/2022/009359.

\noindent\textbf{Author contributions.}\ 
Abhishek contributed conceptualization, methodology, software, investigation, visualization, and preparation of the original draft. Nabeel Salim contributed validation, analysis, and review and editing of the manuscript. P. Arumugam contributed conceptualization, supervision, funding acquisition, and review and editing of the manuscript. All authors reviewed and approved the manuscript.

\noindent\textbf{Data availability.}\ 
The numerical arrays required to reproduce the figures and tables are available from the corresponding author upon reasonable request. The manuscript reports the Hamiltonian, active-space convention, ansatz construction, optimizer settings, and exact-validation procedure required to interpret the data.

\noindent\textbf{Competing interests.}\ 
The authors declare no competing interests.

\bibliography{refs}

\end{document}